# Dual-Polarization Bandwidth-Bridged On-Chip Bandpass Sampling Fourier Transform Spectrometer from Visible to Near-Infrared

Kyoung Min Yoo[1] and Ray T. Chen[1,2,]*

[1]Department of Electrical and Computer Engineering, The University of Texas at Austin, 10100 Burnet Rd. Austin, TX, 78758, USA.

[2]Omega Optics Inc., 8500 Shoal Creek Blvd., Bldg. 4, Suite 200, Austin, TX, 78757, USA.

*Corresponding author. E-mail: chenrt@austin.utexas.edu, yoo_eb@utexas.edu

# ABSTRACT

The on-chip broadband optical spectrometers which cover the entire tissue transparency window ($\lambda = 650 - 1050\ nm$) with high resolution are highly demanded for the miniaturized bio-sensing and bio-imaging applications. Here, we propose a novel type of spatial heterodyne Fourier transform spectrometer (SHFTS) integrated with a sub-wavelength grating coupler (SWGC) for the dual-polarization bandpass sampling on the $Si_3N_4$ platform. Through tuning the coupling angles with different polarization, we experimentally demonstrated the unprecedented broadband spectrum retrieval results with the overall bandwidth coverage of 400 nm, bridging the wavelengths from 650 nm to 1050 nm, with the resolution of 2-5 nm. By applying the bandpass sampling theorem, we circumvented the intrinsic trade-off limitation between the bandwidth and resolution of SHFTS without having an outrageous number of Mach-Zehnder interferometer (MZI) arrays or adding additional active components. The bandpass sampling SHFTS is designed to have linearly unbalanced 32 MZIs with the maximum optical path length difference of 93 $\mu m$ within an overall footprint size of 4.7 mm x 0.65 mm, and the coupling angles of SWGC are varied from 0° to 32° to cover the entire tissue transparency window.

**Keywords** Spectrometer, Fourier transform, Broadband spectrometer, Spatial heterodyne Fourier transform spectrometer, Bandpass sampling, Near infrared, Silicon nitride waveguide, Sub-wavelength grating coupler, Tissue transparency window.

# INTRODUCTION

Owing to unique "fingerprint" signatures of molecular absorption and Raman spectrum of the individual molecules, the optical spectrometer has been known as an indispensable tool to analyze the optical spectrum for a wide range of applications including biological and chemical analysis, medical diagnosis, environmental and planetary monitoring, to name a few. Particularly, near-infrared (NIR) wavelength is beneficial for adoption in mammalian tissues because of its low absorbance, autofluorescence, and lower light-scattering loss from various mammalian bio-cells including hemoglobin, melanin, fat, lipids, water, and else, when compared to the shorter wavelengths[1-3]. Also, a non-invasive in-vivo diffuse optical characterization of human tissues using optical spectroscopy to assess mean absorption and reduced scattering spectra in a NIR tissue transparency window ($\sim 650 - 1050\ nm$) opens a new possibility of monitoring various vital signatures of bio-analytes[2,4,5]. Conventional spectrometers consist of movable mirrors and free-space optic components are typically bulky and expensive. Moreover, it requires sensitive and precise beam alignments, which place constraints on their applications in particular platforms accompanying the environmental fluctuations, such as airborne or handheld devices. However, the development of on-chip spectrometers based on photonic integrated circuits (PICs) has shown great promise in the sense that it can offer low-cost, portable, and robust spectroscopy, along with low-power consumption and high-reliability[6]. In recent decades, a myriad of on-chip spectrometer devices has been demonstrated based on different operating schemes, such as dispersive optics using arrayed waveguide gratings (AWGs)[7-10], echelle grating[11], metasurface elements[12], arrayed narrowband filters[13], computational spectral reconstruction-based systems[14], and Fourier transform spectroscopy (FTS). Specifically, FTS is a technique that measures the spectrum with the interference of light instead of dispersion, offering advantages including high optical throughput and multiplexing advantage, in turn, larger signal-to-noise ratio (SNR) compared to grating-based dispersive counterparts. Several on-chip FTS operation schemes were proposed, including stationary-wave integrated Fourier transform (SWIFT)[15,16] spectrometers, micro-ring resonator (MRR) assisted Fourier-transform (RAFT)[17] spectrometer and spatial heterodyne FTS (SHFTS)[18-21]. In the SWIFT scheme, the interference patterns created from the interference of two adjacent propagating waves through the parallel waveguides are diffracted out-of-plane and monitored by the external detectors such as an array of photodetectors (PDs)[15]. To achieve the full interferogram completely, the pitch size of the PD arrays should be smaller than that of the interference pattern to avoid the subsampling error[15]. Consequently, the resolution and bandwidth of the SWIFT spectrometer are highly limited by the minimum pitch size of the PD array, which puts a significant constraint on practical applications.

The SHFTS consists of an array of unbalanced Mach-Zehnder interferometers (MZIs) with linearly increasing optical path delays (OPDs)[18,19]. In this concept, the output powers of each MZI configure each point of the spatial interferogram which can be captured independently by a linear on-chip PD array, allowing the acquisition of the entire interferogram in a single capture without any moving parts or external analyzers. However, the spectral resolution and bandwidth of SHFTS are closely related to the number of MZIs and maximum OPD, which always requires a balance to meet the limited chip-scale and detecting conditions. In other words, in a standard SHFTS configuration, there is a trade-off between the resolution and bandwidth, hence achieving a fine resolution with broadband operation requires an unrealistically large number of MZI arrays, which increases the size of the device footprint significantly. For example, we previously demonstrated the standard SHFTS device with 24-MZI array for the resolution of 5 nm with the bandwidth of 60 nm centered at $\lambda_o$ =900 nm in the silicon-nitride ($Si_3N_4$) platform[22]. However, to cover the visible (VIS) to near-infrared (NIR) tissue transparency window, it requires broadband operation with ~400 nm bandwidth. To increase the bandwidth while maintaining 5 nm resolution, it requires more than 160 MZI array, which is impractical to be implemented as an on-chip device. To overcome this restriction, several approaches were proposed such as thermally-tuned MZI structures[23,24], RAFT[17], and applying the compressive-sensing scheme[25]. By introducing the thermally tunable MZI and MRR structures, the high-resolution spectrum retrieval results with sub-nanometer resolution and ~100 nm bandwidth from one output channel were reported[17]. In spite of the advantages of the thermo-optic effect contributing to the small-footprint and high-resolution performance, it takes a longer time to collect each point of the thermally induced interferogram separately with corresponding incremental changes of temperature ($\Delta T_{step}$), controlled by the thermoelectric temperature controller (TEC) to achieve the uniform sampling of the thermo-optical induced OPD, which also makes the overall system configuration complex compared to the standard SHFTS.

In this paper, instead of adding additional active components, we propose and experimentally demonstrate an alternative approach using a bandpass sampling theorem with standard SHFTS configuration integrated with a sub-wavelength grating coupler (SWGC) to achieve the overall bandwidth covering the entire NIR tissue transparency window (650 nm to 1050 nm) with < 5 nm resolution. To build the on-chip SHFTS device for NIR wavelength, we chose the $Si_3N_4$ platform due to its low material absorption loss over the wide spectral range from 400 nm up to 2.3 $\mu m$ as well as the high refractive index contrast between the silicon-dioxide ($SiO_2$) cladding ($n_{SiO_2} \approx$ 1.46 @ $\lambda = 900\ nm$) and the $Si_3N_4$ core ($n_{Si_3N_4} \approx 2.01$ @ $\lambda = 900\ nm$).[26] However, the refractive index of $Si_3N_4$ varies depending on the deposition techniques and the quality of films. Typically, low-pressure chemical vapor deposition (LPCVD) deposited $Si_3N_4$ has a higher refractive index than the plasma-enhanced CVD (PECVD), so we determined to use LPCVD deposited $Si_3N_4$, and applied corresponding refractive index values for the device design and simulation[27]. Finally, we have developed and fabricated the first prototype SHFTS chip on a CMOS-compatible $Si_3N_4$ platform and demonstrated experimental results.

## RESULTS

### Concept and principle

The theory and principle of the standard SHFTS have been demonstrated by Florjańczyk et al.[18] previously. As briefly mentioned in the introduction, SHFTS consists of an array of unbalanced MZI with linearly increasing OPD with constant increment across the array configuring spatial interferogram. For a given single-input source, the phase change from each MZI is converted into an intensity change based on interferometric schemes. The input spectrum can be retrieved through the discrete Fourier transform (DFT) which can be written as[18]

$$p^{in}(\bar{\sigma}) = \frac{\Delta x}{N} P^{in} + 2 \frac{\Delta x}{N} \sum_{i=1}^{N} F(x_i) cos 2\pi \bar{\sigma} x_i, \text{ where } F(x_i) = \frac{1}{B_s}(2P_i^{out} - A_s P^{in}) \dots (1)$$

Here, $P^{in}$ is the input power, N is the number of MZIs, $\sigma$ is the wavenumber, and $\bar{\sigma} = \sigma - \sigma_{min}$ is the shifted wavenumber, where $\sigma_{min}$ represents the minimum wavenumber at the Littrow condition; at the Littrow condition, the phase delays in different MZIs are integer multiplies of $2\pi$, so the output powers of each MZI ($P_i^{out}$) are constant. $\Delta x$ is the maximum interferometric delay, that is $\Delta x = n_{eff} \Delta L_{max}$, where $n_{eff}$ is the effective index of the strip waveguide and $\Delta L_{max}$ is the maximum path delay of the most unbalanced MZI. The spatial interferogram $F(x_i)$ is discretized at N equally spaced OPD values $x_i$ ($0 \leq x_i \leq \Delta x$) which is defined as $x_i = n_{eff} \Delta L_i$, where $\Delta L_i$ is the path length difference of the $i$ th unbalanced MZI. The input power $P^{in}$ is constant for all the MZIs, and $P_i^{out}$ represents the output power of the $i$ th MZI with the coupling and loss coefficients of the MZI components $A_s$ and $B_s$. As the wavenumber of monochromatic input $\sigma$ changes from the Littrow wavenumber, $P_i^{out}$ distribution becomes periodic, and different wavenumbers create different periodic patterns. Subsequently, a polychromatic input signal, which can be considered as a superposition of monochromatic constituents, creates a corresponding spatial interferogram pattern formed by a superposition of the respective periodic $P_i^{out}$ fringes from monochromatic input. The resolution of spectrometers, represented by the wavenumber resolution $\delta\sigma$ is determined by the maximum interferometric delay $\Delta x$. To resolve two monochromatic signals separated by $\delta\sigma$ at the most unblanced MZI with OPD of $\Delta x$, the phase change from respective interferograms of $\sigma$ and $\sigma + \delta\sigma$ should be differ by one fringe ($2\pi$), that is $\Delta\varphi = 2\pi(\sigma + \delta\sigma)\Delta x - 2\pi\sigma\Delta x = 2\pi$, in turn,[18]

$$\delta\sigma\Delta x = \delta\sigma\Delta L_{max} n_{eff} = 1 \ldots (2)$$

Thus, the maximum path delay of the MZI array ($\Delta L_{max}$) can be designated as follows:

$$\Delta L_{max} = \frac{1}{\delta\sigma \cdot n_{eff}}, where\ \delta\sigma = \frac{1}{\lambda_o} - \frac{1}{\lambda_o + \delta\lambda} \ldots (3)$$

where $\lambda_o$ is the center wavelength and $\delta\lambda$ is the wavelength resolution. The number of MZIs in the array (N) is equivalent to the discrete sampling points in the spatial interferogram. Based on the Nyquist-Shannon sampling theorem, the minimum sampling points ($N_{min}$) required to fully reconstruct the input spectrum ($p^{in}(\bar{\sigma})$) within the band-limit range is determined as follows:

$$N_{min} = 2\Delta x\Delta\sigma = 2\frac{\Delta\sigma}{\delta\sigma} = 2\frac{\Delta\lambda}{\delta\lambda} \cdots (4)$$

where $\Delta\sigma$ and $\Delta\lambda$ are the wavenumber and wavelength bandwidth of the spectrometer. Based on the sampling thoery, the wavenumbers that are equally distributed above and below the Littrow wavenumbers produce the same interference fringe patterns, and the input spectrum reconstructed by the DFT of spatial interferogram creates the wavenumber-shifted replicas of the original transform $p^{in}(\bar{\sigma})$ above and below the band-limit as shown in the Fig. S1a. Consequently, when the input spectrum contains the signals outside of the band-limit exceeding the bandwidth, the reconstruction aliasing error due to the overlap of these copies (flipped images) makes the retrieved spectrum indistinguishable, as described in Fig. S1b, and this intrinsic constraint puts a limitation to retrieve the broadband continuum input signals from the Fourier transform system.

In signal processing, a technique which is known as bandpass sampling or undersampling has been used to reconstruct the signal with sampling rate below the minimum Nyquist rate (Eq. 4) using bandpass filter[28,29]. In this paper, we exploit this concept into the SHFTS configuration to solve the trade-off condition between the bandwidth and resolution by employing the bandpass filter to divide the broadband-continuous spectrum into multiple narrow-band channels. Thereby, it is able to reconstruct each band without aliasing error, rather than retriving the signals all together in a single-band. In consequence, we were able to acheive the broad overall bandwidth coverage ($\Delta\lambda_{overall}$) without degrading the resolution.

The grating coupler (GC) is one of the essential components to build PIC chip for the fiber-to-chip coupling[30,31]. In general, the narrow-bandwidth operation, polarization (transverse-electric: TE and transverse-magnetic: TM) and angular-dependencies of GC are considered as intrinsic disadvantages to be employed as the broadband coupler. However, these properties are very beneficial for the bandpass-sampling scheme, since it can be implemented as a coupler as well as a polarization-selective bandpass filter; in addition, the coupling wavelength can be shifted by the coupling-angle ($\theta$) tuning based on the phase-matching condition[32], which makes it a tunable-bandpass filter.

Fig. 1 demonstrates the overall design and operation principle of a $Si_3N_4$ bandpass sampling SHFTS integrated with SWGC which covers the entire NIR tissue transparency window. The SHFTS consists of an array of 32 MZIs with single SWGC input port (Fig. 1a). The broadband input signal is coupled from the fiber to the waveguide through the SWGC (Fig. 1d), and the input spectrum is bandpass-filtered based on the SWGC's phase-matching condition, which can be tuned by the coupling angle ($\theta$) as shown in Fig. 1e. Namely, the broadband input spectrum is bandpass sampled into multiple narrow-bands by different coupling angle conditions, which completely separates the aliases (anti-aliasing). The coupled light is equally divided into linearly unbalanced 32 MZIs through the cascaded Y-splitter, and the intensities of each MZIs output powers are measured, forming the spatial interferogram (Fig. 1b). And then, each sampled input spectrum can be fully recovered using eq. (1) without overlapped aliases, and finally the overall broadband spectrum is retrieved by the superposition of separated bands (Fig. 1c).

To build the on-chip SHFTS with the low-loss operation in a NIR tissue transparency window ($\sim 650 - 1050\ nm$), we designed and optimized the $Si_3N_4$ passive components including the strip waveguide, Y-branch splitter and combiner, and SWGC using the Lumerical simulation tool. In the design of the prototype, we used the SWGCs for the MZI output ports to measure the output powers, which can be further integrated with a PD array by virtue of the modern PIC technology[33], providing much more compact and stable devices. In the following sections, we will demonstrate the simulation and experimental results of the optimized passive components and the first prototype of the SHFTS device. The optimal design and simulation results of $Si_3N_4$ strip waveguide (Fig. S2) and Y-branches (Fig. S5) were described in the supplementary.

**Sub-wavelength grating coupler (SWGC)**

A conventional GC with shallow etched gratings requires additional alignment lithography steps, which make the whole fabrication process complex. To alleviate the fabrication complexity, several through-etched GC designs using sub-wavelength grating (SWG) structures to reduce the Fresnel back reflection has been reported in various platforms[34,35]. In this work, by implementing the SWG structures between the major Si₃N₄ gratings, it was able to get a through etched structure which can be patterned and etched altogether with other passive components without additional alignment patterning steps.

Also, it is known that the central coupling wavelength $\lambda_o$ of the grating coupler is determined as following equation based on the phase matching condition[32].

$$\lambda_o = (n_{eff} - n_c sin\theta)\Lambda \cdots (5)$$

where $n_{eff}$ is the effective index of the grating, $n_c$ is the refractive index of the cladding, $\theta$ is the coupling angle and $\Lambda$ is the grating period. Therefore, the coupling wavelength $\lambda_o$ can be shifted by $\theta$ tuning based on the eq. (5). Here, we designed the Si₃N₄ SWGC integrated with SHFTS chip, serving not only as a coupler but also a tunable bandpass filter for the bandpass sampling of the broadband input signals. The schematic illustration of the SWGC is shown in Fig. 2a and b.

A simulation model (Fig. 2c) has been created using 2D and 3D finite-difference time-domain (FDTD) simulation to optimize the device for the fundamental TE mode with small insertion loss as well as small back reflection to the waveguide. Design parameters including the grating period ($\Lambda$), the width of the grating ($w_g$), and the width of the sub-wavelength grating ($w_{swg}$) were optimized with the built-in particle swarm algorithm in the simulation tool. Furthermore, the thickness of the SiO₂ bottom cladding was optimized as $h_{core}$ = 2.8 $\mu m$ to get the maximum coupling efficiency as shown in Fig. S3a. The refractive index profile and E-field of the optimized SWGC structure are shown in Fig. S3b and Fig. 2d, showing that the light is coupled from the fiber to the waveguide through the SWG structures. The video animations of the E-field simulations with different coupling angles can be found in the supporting information. Fig. 2e shows the coupling efficiency of the optimized SWGC with $\theta = 11°$ for both TE and TM modes as a function of the wavelength, extracted from the S-parameters of the input and output ports. The coupling efficiency unit is normalized as maximum of 1, indicating that 100% of light is coupled from the fiber to the waveguide, and the reciprocal S-parameters indicated that the coupling efficiency from the fiber to the waveguide is identical with the opposite direction. Moreover, we can see the coupling selectivity between the TE mode and TM mode, and the SWGC has ~3 dB coupling loss with TE mode at $\lambda_o = 900$ nm, along with the 3 dB bandwidth of ~50 nm. Finally, we swept the coupling angle from $\theta = 0°$ to 31° and monitored the $\lambda_o$ shifts from 1030 nm to 650 nm, and the results in XYZ plot and XY plot are shown in Fig. 2f and Fig. S4. The maximum coupling efficiency of ~65% was achieved at $\lambda_o = 650$ nm with θ = 29°, and the center wavelength shifts to the longer wavelength up to 1030 nm as the coupling angle decreases to surface normal. Based on the optimal design and simulation results, we fabricated the device and inspected the dimension deviation by scanning electron microscope (SEM) due to the fabrication error as in Table 1, and the corresponding SEM image can be found in Fig. S8.

Since these deviations of the device dimensions are critical to the phase matching condition of SWGC in turn the coupling efficiencies and wavelengths, we carefully examined the simulation results with the real-fabricated dimensions in Table 1 again, that is shown in Fig. 2g. Comparing with the ideal simulation results in Fig. 2f, we can see that the overall spectrum shifts to the longer wavelength slightly. Even though the overall spectrum has red shift due to the fabrication error, we could examine the wavelength shift from 650 to 1050 nm by tuning the coupling angle. However, we found that the SWGC cannot effectively couple some of the wavelength ranges with TE mode, especially between the $\lambda \cong 700$ nm and 800 nm. We presumed that this is due to the vertical-destructive inteference between the $Si_3N_4$ and $SiO_2$ layers. To address this issue and couple the entire wavelength range, we utilized the TM mode which has different phase matching condition from the TE mode. Fig. 2h shows the TM mode simulation results using the same SWGC structures. Although the SWGC structure is optimized for the TE mode, the SWGC can couple the TM mode even better than TE mode especially for the shorter wavelenghs, which allows bridging the TE mode forbidden wavelengths between 700 nm and 800 nm. Accordingly, the entire wavelenth from 650 nm to 1050 nm can be bandpass filtered and coupled into the SHFTS with corresponding polarization and coupling angle (θ) conditions.

**Bandpass sampling SHFTS design**

Using the optimized $Si_3N_4$ components described above, we designed and analyzed the bandpass sampling SHFTS. Based on the Eq. (3) and (4), the number of the MZI array (N) and the maximum OPD ($\Delta L_{max}$) are designed to specifiy the bandwidth (Δλ) and the resolution (δλ) of the SHFTS; the bandwidth of SHFTS is designed to be wider than that of the SWGC, preventing the overlap of sampled aliases. As shown in Fig. 2f and g, the 3dB bandwidths of SWGC increases from ~30 nm to 70 nm as the $\lambda_o$ increases, therefore the bandwidth of SHFTS should be designed to be wider than that. In other words, the bandwidths of each narrow-band channels of SHFTS must be able to fully cover the sampled signals which are coupled from the SWGC. Here, we designed SHFTS with N=32 and $\Delta L_{max}$ = 93 $\mu m$ (Fig. 3a) to have δλ < 5 nm, and Δλ = 30 ~ 80 nm at the wavelength range of λ=650 ~ 1050 nm. To examine the resolution and bandwidth of SHFTS, we built the interconnect simulation model of SHFTS using the optimized components as shown in Fig. 3b and c, and the $\lambda_o$ of narrow-band input signals were swept from 591 nm to 1060 nm to get the ranges of the narrow-band channels and corresponding resolutions.

The spatial interferogram of output powers from 32 MZIs were measured by optical powermeter (Fig. 3d) and the input signals were retrieved by using Eq. (1) in MATLAB as shown in Fig. 3e. Then, we were able to reconstruct the whole broadband signals by integrating each narrow-band channels by superposition as shown in Fig. 1c. The overall configurations of the narrow-band channels including the corresponding bandwidths and resolutions are shown in the Table 2.

Finally, in the experiment results described in the next section, we demonstrated the fabrication and measurement results in detail.

**Experimental results**

By virtue of a through-etched SWGC design, We were able to pattern and etch the whole device in one lithography and etching step without any additional alignment lithography process. The final footprints of the fabricated chip including the loss characterization patterns and bandpass sampling SHFTS device are shown in Fig. S7. The optical microscope and SEM images of fabricated devices are shown in Fig. 4. The fabrication results show a clean integrity without stitched or collapsed structures.

Then, we built the measurement setup to test the fabricated device. The schematic diagram and real picture of the measurement setup are shown in Fig. S9a and b. We used the broadband supercontinuum light source (NKT photonics SuperK laser) to examine the performances of the fabricated SWGC and bandpass sampling SHFTS. First of all, to verify that the broadband light is coupled to the PM-SMF succefully, we measured the fiber coupled SuperK laser's spectrum by the optical spectrum analyzer (OSA) directly. Fig. S10 shows the pictures of the superK laser coupling setup and the OSA measured optical spectrum from the fiber, and we examined that the broadband light is coupled to the PM-SMF successfully with the maximum total optical power of ~130 mW. Using the measurement setup and the loss characterization devices consist of different lengths of waveguides, the propagation loss of the $Si_3N_4$ strip waveguides and the loss from the Y-branch were measured as ~2.5 dB/cm at $\lambda$=905 nm by the cut-back method (Fig. S11), and 0.5 dB loss from the Y-splitter.

And then, the output spectrum of the SWGC were measured with different coupling angles to test the coupling efficiencies and the wavelength shifting. The input and output coupling angles were equally controlled from 0° ~ 33° as shown in the pictures in Fig. 5a, and the output fiber was connected to the OSA to measure the spectrum directly.

Fig. 5b and c show the SWGC spectrum measurement results by the OSA with TE and TM modes. With the fundamental TE mode (Fig. 5b), the SWGC shows 3~4 dB losses, and the center wavelength $\lambda_o$ shifts from 650 nm to 1000 nm as the coupling angle changes from 33° to 0°, with the corresponding 3dB bandwidth changes from 30 nm to 70 nm. The TM mode coupling efficiencies with two different coupling angles $\theta = 18°\ and\ 12°$ were also measured for $\lambda_o \cong 740$ nm and 835 nm as shown in Fig. 5c. These experimental results were well matched with the simulation results in Fig. 2g and h, respectively. Based on these results, we verified the coupling wavelength shifting range and efficiencies of the SWGC experimentally, covering the entire tissue transparency window ($\lambda = 650 - 1050\ nm$). As a consequence, the superK broadband input source (Fig. S10c) can be bandpass filtered and divided into the multiple narrow-band channels shown in Table 2 through the SWGC and coupled into the waveguide.

At last, using the same measurement setup, we coupled the SuperK laser light into the bandpass sampling SHFTS and measured the output powers of each MZI's interferogram using different SWGC coupling angles to reconstruct the input spectrum. Single input signal is equally divided into 32 MZI channels by the cascaded Y-splitters, and the output powers of each MZI are measured separately by the optical powermeter. The optical images of the light-coupled SHFTS chip can be found in Fig. S12. We've measured the optical powers using 8 different coupling conditions, which are $\theta = 35°, 25°, 20°, 14°, 4°, and\ 0°$ with TE polarization, and $\theta = 18°\ and\ 12°$ with TM polarization to retrieve the entire tissue transparency wavelength range from 650 nm to 1050 nm.

The experimental output power measurement results from 32 MZIs using different SWGC coupling conditions are shown in Fig. S13. After we collected the output powers from each MZI, we reconstructed the optical spectrum of the bandpass-sampled signals by the MATLAB code based on DFT equation (Eq. 1), and the results are shown in Fig. 6. To validate the spectrum retrieval accuracy, the FTS retrieved results (red-solid lines) were compared with the direct OSA measurement results (black-dotted lines). The retrieved spectra are well matched with the direct OSA measurement results but the discrepancies are mainly due to the sampling optical path delay (OPD) errors induced by the inevitable fabrication deviations and detecting noise from the optical powermeter. Fig. S14 shows the overall retrieved spectra from $\lambda = 650 - 1050\ nm$ by the superposition. In conclusion, we experimentally retrieved the input spectrum from $\lambda = 650\ nm\ to\ 1050\ nm$ with 8 discrete narrow-band channels by the bandpass sampling SHFTS through SWGC coupling.

## DISCUSSION

Our proof-of-concept experiment demonstrates the broadband spectrum retrieval performance using the broadband supercontinuum light source and bandpass sampling SHFTS, representing the overall spectrometer bandwidth coverage ($\Delta\lambda_{overall}$) of 400 nm without compromising the spectrometer resolution. To the best of our knowledge, this is the first demonstration of an on-chip broadband FTS covering the entire tissue transparency window without adding additional active components for electrical or thermal enhancements. Previously reported miniaturized spectrometers at VIS-NIR wavelength ranges (Table 3) were either limited to the narrow bandwidth coverage or poor resolution. Or the measurement system requires too large and complicated configuration[11] which cannot be implemented as handheld or portable applications. In comparison to these spectrometers including thermally tunable FTS devices, the bandpass sampling FTS concept stands out in the sense that it provides broadband coverage while maintaining a fine resolution and relatively small size of the device footprint, without external active components to introduce additional thermal or electrical induced OPDs. Moreover, the bandpass wavelength ranges are determined by the coupling angle of SWGC and each discrete channel can be retrieved in a single capture with a fixed coupling angle, which is beneficial to the real-time biosensing applications especially in the tissue transparency wavelength window.

Despite the advantages of the bandpass sampling SHFTS described above, our prototype result leaves much room for further improvement for the monolithic integrated circuits and chances for the sensing applications, discussed in the following sections.

**On-chip broadband light source and photodetector integration**

In the broadband spectrometer experiments presented in this work, we used the benchtop supercontinuum laser source with the objective lens to couple the light into the single-mode fiber. However, to realize the miniaturization of the whole FTS system, the integration of the on-chip broadband light source and photodetector arrays is required. Several broadband LED sources are available in the market, providing the spectral emission from $\lambda = 400\sim1000\ nm$; however, using the external LED sources is not an attractive solution to realize the on-chip spectrometer device ultimately, because of the high coupling losses and packaging challenges and costs involved. Various researches have been conducted for the efficient PIC integrated light sources[37], including the heterogeneously integrated on-chip optically-pumped LED sources based on the InP platform[38], colloidal quantum dot integration on $Si_3N_4$ platform[39], on-chip supercontinuum sources[40], and else. Especially, Haolan et al.[40] reported the generation of an octave-spanning supercontinuum covering $\lambda = 488 - 978\ nm$ using a 1-cm-long $Si_3N_4$ waveguide with Ti:Sapphire pump laser at 795 nm. These advancements show a great promise for the monolithic integration of the broadband light source with the waveguide-based spectrometer device on the $Si_3N_4$ platform. Moreoever, to exploit the on-chip integrated broadband light source into the bandpass sampling SHFTS scheme, a proper method that can substitute the SWGC in this work should be devised as an on-chip tunable bandpass filter. A recent research reported the microelectromechanical (MEMS)-based tunable grating coupler[41], which is applicable in our SWGC design to actively control the coupling angle without moving the fiber or beam angle. Moreover, the other research proposed the single-channel optical bandpass filter based on plasmonic nanocavities[42], showing a single transmission peak over a wavelength range from 400 to 2000 nm. Hence, we expect there is a decent potential towards the PICs integrated bandpass sampling SHFTS with the on-chip broadband light sources. In addition, a lot of researches have been presented the vertically- or end-coupled integrated photodetectors using various platforms such as germanium (Ge) on SOI[33], graphene[43], or other advanced 2D materials[44]. For the meaningful benchmarking, we focused on the CMOS compatible photodetectors, especially for the arrayed waveguide integrated applications. Recently, Hongqiang et al.[33] reported the vertically coupled Ge PD array integrated with silicon waveguides for the on-chip arrayed waveguide grating (AWG) interrogator with around 90 % photon absorption and quantum efficiencies. The same concept can be used at the output ports of the MZI array in our SHFTS device to measure the output powers of each MZI interferogram as shown in the conceptual diagram in Fig. 1a, but the sensitivity and the noise level of the PD array should be explained experimentally to ensure the spectrometer performance. Based on these estimations, we expect further improvements of the compact, cost-effective, and reliable on-chip VIS-NIR spectrometer device based on the bandpass sampling SHFTS integrated with the on-chip broadband light source and PD array in a single chip, which enable the portable handheld spectrometer devices for the real-time in-vivo bio-photonic applications.

**Towards compact portable spectrometers for biosensing applications**

The most essential value of a NIR tissue transparency window spectroscopy is that the light can penetrate into the skin beyond several millimeters of tissue benefit from the reduced tissue scattering and autofluorescence in these spectral ranges[2,3], therefore the optical characteristics of various endogenous and exogenous tissue species can be measured based on the non-invasive assessment of the Raman scattering[4] or the absorption spectra to detect the glucose[45,46], tissue oxygen saturation level[5,47], and the skin-cancer detection[48], and various skin conditions[49]. However, the critical issues of the spectrometer system that require further elaboration for the practical bio-sensing applications are the power and loss characterizations. Considering the entire device configuration for the in-vivo non-invasive biosensing applications which would be applied on top of the skin with fiber-optic probe[48,50], a thorough investigation of the loss from the skin and device operation is required to assess how much input power should be utilized to reconstruct the signals. According to the recent study reported by Tianxing et al.[45], when linearly polarized NIR light illuminates the skin surface, approximately 58% (3.77 dB loss) of the light cannot interact with the tissues below the skin due to the absorption (53%) and direct reflection from the skin surface (5%), and the rest of the light penetrates the skin and may interact with tissues before and while being reflected back, but the majority of them (> 95%) will be depolarized. Since the SWGC in this work is polarization-dependent, the linear polarizer is required to filter TE or TM mode before the SWGC stage, which brings an additional ~3 dB loss. In total, there will be an inevitable ~7 dB loss from the skin approximately. Then, taking into account the experimentally measured loss from the input SWGC (~4 dB), cascaded Y-branches and waveguide propagation loss (~4 dB), we will have around 8 dB loss from the SHFTS device operation. Here, assuming the input source has the peak power of 10 dBm (10 mW), around -20 dBm (10 µW) can be transferred to each output photodetector in the array after being divided into 32 MZIs. However, we expect to have an additional loss from the skin to fiber-optic probe coupling, which must be further demonstrated experimentally. Taking all of this into account, we believe that this research constitutes the first step to realize fully integrated portable bio-spectrometer sensors.

**Summary**

In this paper, we designed and experimentally demonstrated the bandpass sampling SHFTS integrated with SWGC on the Si$_3$N$_4$ platform to achieve broad bandwidth coverage ($\Delta\lambda_{overall} = 400\ nm$) without compromising the spectrometer resolution ($\delta\lambda \approx 2 - 5\ nm$) in a VIS-NIR tissue transparency window ($\lambda = 650 - 1050\ nm$). Unlike the standard SHFTS operating scheme, the bandpass sampling theorem is applied by tuning the coupling angle of SWGC, substituting the tunable-bandpass filter in our design. Namely, the SWGCs were used not only as a fiber to waveguide coupler but also as an anti-aliasing filter, dividing the continuous broadband spectrum into multiple narrow-band channels. Then, the original broadband spectrum can be reconstructed by the superposition of each retrieved narrow-band spectra. We optimized the low-loss passive components to build the MZI structures, including the strip waveguide, Y-branch, and SWGC, and the SHFTS is designed to have linearly unbalanced 32 MZIs with the maximum OPD ($\Delta L_{max}$) of 93 $\mu m$, with the total footprint size of ~4.7 mm x 0.65 mm. To the best of our knowledge, this is the first experimental demonstration of an on-chip broadband FTS covering the entire tissue transparency window without adding additional active components. We also discussed the possibilities and challenges of the integration with the on-chip broadband light source and PD array, which provide the guidelines to further improve the performances towards the fully integrated portable spectrometers for the lab-on-a-chip Raman or absorption spectroscopy systems.

**METHODS**

**Device simulation and optimization**

Simulation software Lumerical MODE and FDTD (finite-difference time-domain) were used to analyze and optimize the passive components. The internal optimization tool with particle-swarm-algorithm was used to generate a number of structures with different dimensions in specific ranges and get the maximum coupling efficiency or minimum losses as needed. Also, the S-parameters from the input and output ports of SWGC and Y-branches were calculated from the FDTD simulation to monitor the light transmission characteristics, and the S-parameters from the optimized passive components were used to build the SHFTS model in the interconnect photonic circuit simulation.

**$Si_3N_4$ SHFTS device fabrication**

The $Si_3N_4$-on-$SiO_2$ wafers were prepared with a 220 nm thick LPCVD grown $Si_3N_4$ on a 2.8 $\mu m$ thick $SiO_2$ bottom cladding on a silicon substrate (Fig. S6a) from Rogue Valley Microdevices Inc.. Then, ~400 nm thick Ebeam resist (ZEP-520A) is deposited on top of $Si_3N_4$ layer by spin-coating. The patterning is done by JEOL E-beam (JBX-6000FS) lithography tool, followed by developing in n-Amyl acetate for 2 min, and rinsing in isopropyl alcohol (IPA) (Fig. S6b). Following this, the pattern is transferred to $Si_3N_4$ layer by reactive-ion-etching (RIE) (Fig. S6c). Finally, the remaining resist and polymers are cleaned using removal PG and followed with cycles of Acetone/IPA post process treatment (Fig. S6d).

**Measurement setup**

The unpolarized broadband supercontinuum light source (NKT photonics SuperK Versa), which covers the wavelength range from 550 nm to 1750 nm, is coupled to the polarization-maintaining single-mode fiber (Thorlab PM780-HP, PM630-HP) using Glan-Thompson linear polarizer (Newport 5525) and microscope objective lens (100x) for TE or TM mode transmission. The fibers are mounted on the goniometer stages to control the coupling angles for both input and output fibers. The device is placed on the flat stage and the input/output fibers are aligned with the SWGCs to couple the light to the waveguide, monitored by the optical microscope. Finally, the output fiber is connected to the optical powermeter (Newport 2936-C) to measure the output powers.

**Acknowledgements**

This research was supported by the Air Force Research Laboratory (AFRL) Contract # FA864920P0971 and NSF Award #1932753. The authors acknowledge Dr. Xiaochuan Xu and Hamed Dalir's contribution on the initial idea developing, and also wish to thank the anonymous reviewers for their valuable comments and suggestions.

**Author Contributions**

The main idea was conceived by K.M.Y. and R.T.C.. K.M.Y. conducted the device optimizations, simulations, and experiments. K.M.Y. fabricated the device in Microelectronics Research Center at UT Austin, and built the measurement setup and collected the experimental data. K.M.Y. and R.T.C. analyzed and discussed the data and K.M.Y. wrote the manuscript. R.T.C. and K.M.Y. revised and finalized the manuscript.

**Conflict of Interest**

The authors declare no competing interests.

**Supporting Information Available:**

The following files are available free of charge.

Supporting Information (PDF).

Video1: E-field animation of the SWGC fiber to waveguide coupling with $\theta = 3°$.

Video2: E-field animation of the SWGC fiber to waveguide coupling with $\theta = 11°$.

Video3: E-field animation of the Y-splitter at $\lambda = 900\ nm$.

## Tables and Figures

|  | Design | Fabrication | Deviation |
|---|---|---|---|
| **Si$_3$N$_4$ thickness (h$_{core}$)** | 220 nm | 228.6 nm | + 8.5 nm |
| **SiO$_2$ thickness (h$_{clad}$)** | 2.8 um | 2.813 um | + 13 nm |
| **Grating period (Λ)** | 706 nm | 708.5 nm | + 2.5 nm |
| **Grating width (w$_g$)** | 430 nm | 433 nm | + 3 nm |
| **SWG width (w$_{swg}$)** | 65 nm | 85.6 nm | + 25.6 nm |

**Table 1.** SWGC dimension characterization

| Band | Start wavelength [nm] | End wavelength [nm] | Bandwidth [nm] | Resolution [nm] |
|---|---|---|---|---|
| 1 | 647 | 680 | 33 | 2.06 |
| 2 | 680 | 715 | 35 | 2.19 |
| 3 | 715 | 755 | 40 | 2.50 |
| 4 | 755 | 800 | 45 | 2.81 |
| 5 | 800 | 855 | 55 | 3.44 |
| 6 | 855 | 915 | 60 | 3.75 |
| 7 | 915 | 985 | 70 | 4.38 |
| 8 | 985 | 1060 | 75 | 4.69 |

**Table 2.** Discrete narrow-band channels of bandpass sampling SHFTS

| Schemes | Principles | Platforms | Resolution [nm] | Bandwidth [nm] | Range [nm] | No. Channel | Footprint [mm$^2$] |
|---|---|---|---|---|---|---|---|
| Dispersive optics | Waveguide grating diffraction[36] | SiON | 40 | 800 | 300-1100 | - | 35 |
|  | MRR assisted AWG[7] | SiN | 0.75 | 52.5 | 831-883 | 70 | 1.44 |
|  | AWG[8] | SiN | 1.5 | 60 | 830-890 | 40 | 0.62 |
|  | AWG[9] | SiN | 0.5 | 4 | 757-762 | 8 | 2.8 |
|  | AWG[10] | SiON | 5.5 | 215 | 740-955 | 41 | 340 |
|  | Planar echelle gratings & single photon detection[11] | SiN/NbN[a] | 7 | 1400 | 600-2000 | 200 | ~30 |
|  | Folded metasurface[12] | SiO2/Si/Au | 1.2 | 100 | 760-860 | - | 7 |
| Narrowband filters | Linear variable optical filter[13] | Multilayered films (SiO2/TiO2) | 2.2 | 170 | 570-740 | - | 100 |
| Reconstructive | PC slabs[14] | SOS | 1 | 200 | 550-750 | 36 | 0.044 |
| Fourier transform | SWIFT[15] | SiN | 6 | 100 | 800-900 | - | 0.1 |
|  | Bandpass sampling SHFTS (this work) | SiN | 2-5 | 400 | 650-1050 | 32 | 3 |

[a] Cryogenic low-noise amplifiers are required for the measurement

**Table 3.** Comparison of miniaturized spectrometers at VIS-NIR wavelength range (650-1050 nm).

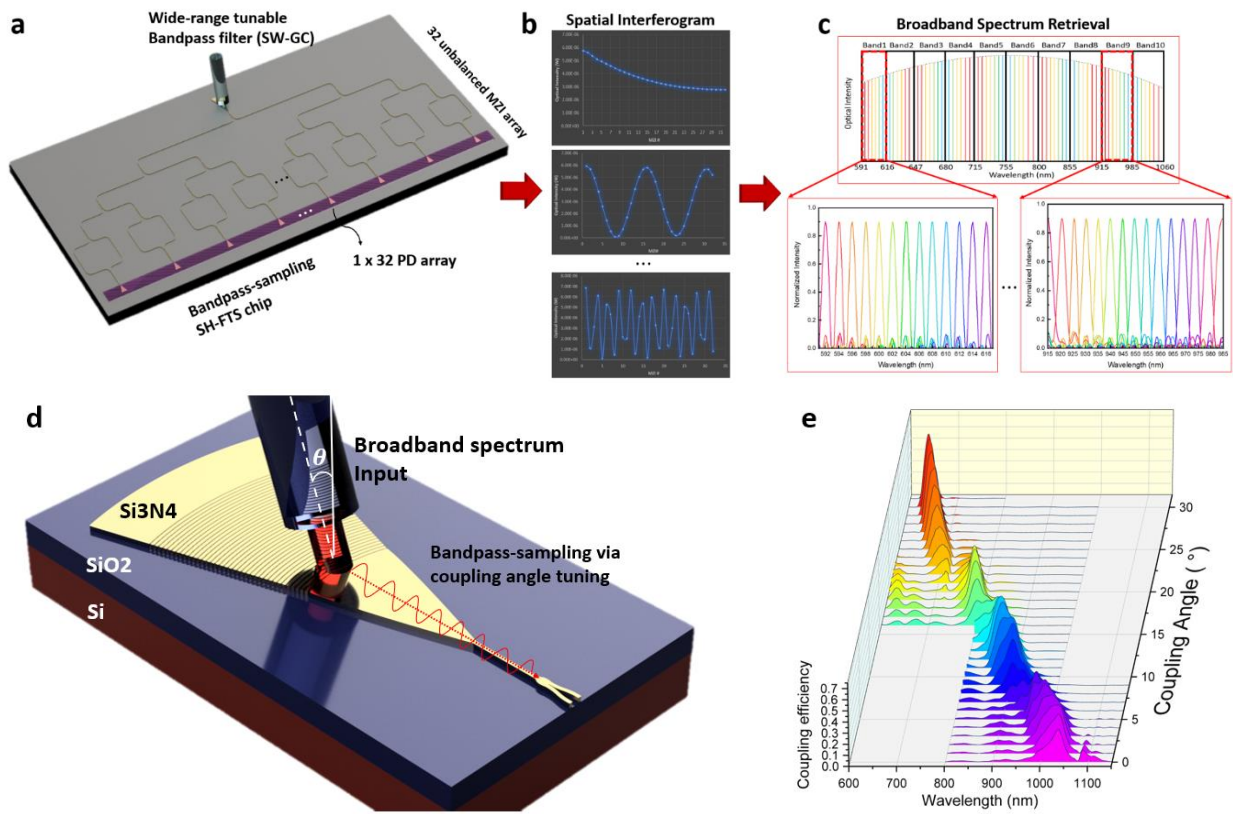

**Figure 1.** Schematic illustration of the bandpass sampling SHFTS and SWGC. **a** The Si$_3$N$_4$ SHFTS chip with an array of 32 MZIs integrated with SWGC and 1x32 PD array; the SWGC is used to couple the light into the waveguide as well as filter the broadband spectrum into multiple narrow-

band channels by coupling angle tuning as shown in Fig. 1d and e. **b** The spatial interferogram consists of output powers from each MZI. **c** The sampled input spectra are retrieved using DFT equation, and the overall spectrum can be reconstructed by the superposition; each separated bands are retrieved with corresponding bandpass sampling conditions from the SWGC, avoiding the aliasing error. **d** the schematic illustration of the SWGC operation with **e** the simulation result showing the coupling wavelength shifting by tuning the coupling angle from 0 to 31 °.

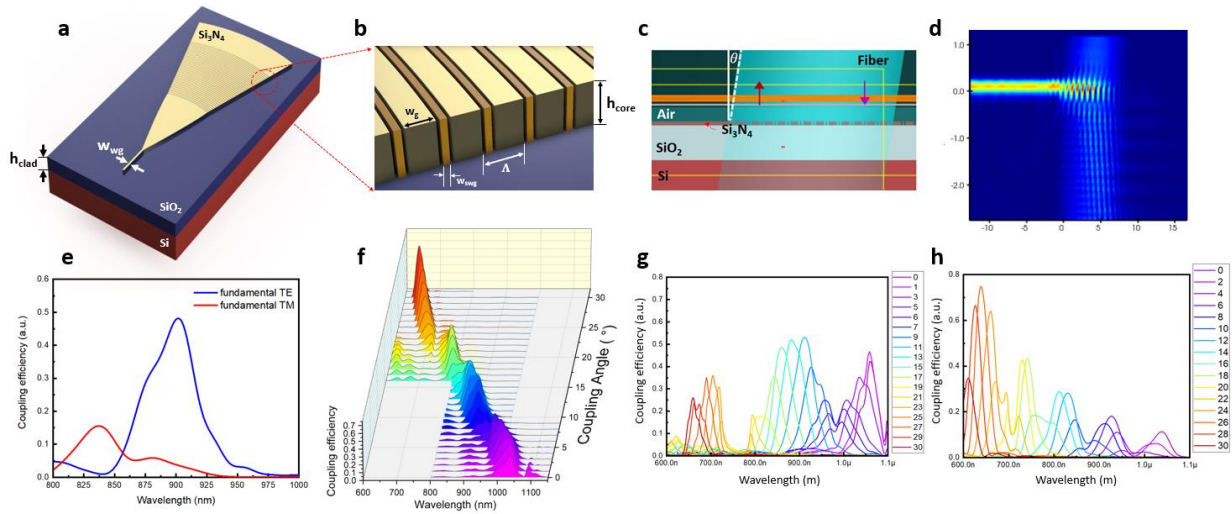

**Figure 2.** SWGC design and simulation results. **a** 3D Schematic illustration of $Si_3N_4$ SWGC and **b** zoomed-in image of the grating structures. **c** The profile of the FDTD simulation model including the SWGC structures and the fiber. **d** E-field profile of the optimized SWGC for TE mode. **e** Coupling efficiencies (S-parameters) of the optimized SWGC with $\theta = 11°$ for the fundamental TE and TM modes as a function of the wavelength. **f** XYZ plot showing the simulation result of the SWGC TE mode coupling wavelength ($\lambda_o$) shift by tuning the coupling angle from 0° to 31°; X-axis: wavelength, Y-axis: coupling efficiency, Z-axis: coupling angle. **g** XY plot of SWGC TE mode coupling simulation result using real (fabricated) dimension, and **h** TM mode coupling efficiencies.

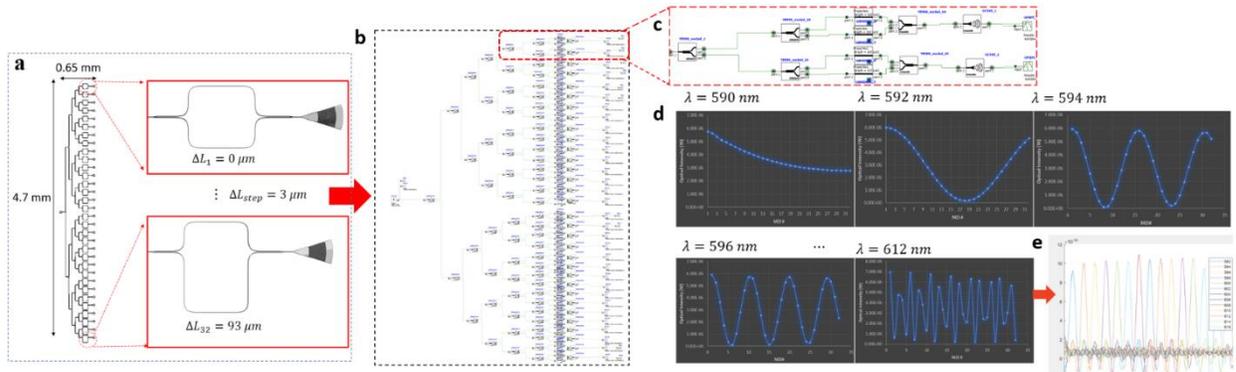

**Figure 3.** SHFTS comprises of 32 MZI array integrated with SWGC input and outputs. **a** Final outlook of the SHFTS device. The overall size is around 4.7 mm x 0.65 mm. **b** Interconnect simulation model using the optimized passive components' S-parameters. **c** Zoomed-in image of the first MZI simulation model. **d** Interferometric patterns of output powers from 32 MZIs with different $\lambda_o$. **e** retrieved input signals by DFT equation using MATLAB data processing.

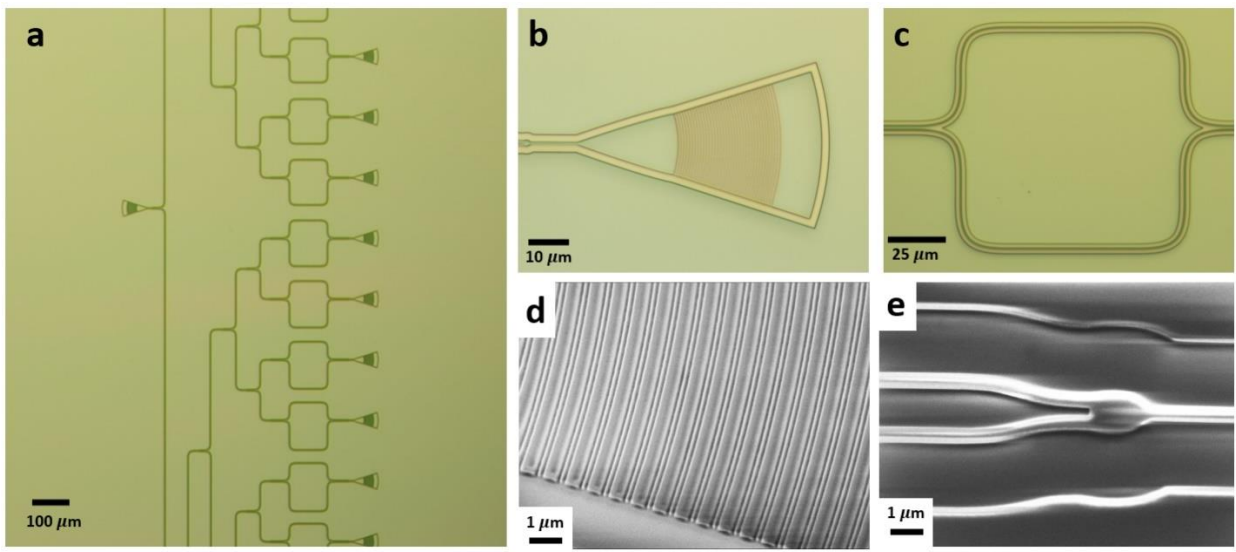

**Figure 4.** The SEM and optical microscope images of the fabricated devices. **a** The optical microscope image of SHFTS, **b** SWGC connected with Y-branch and waveguides, **c** imbalanced MZI, and **d** SEM images of SWGC, and **e** Y-branch.

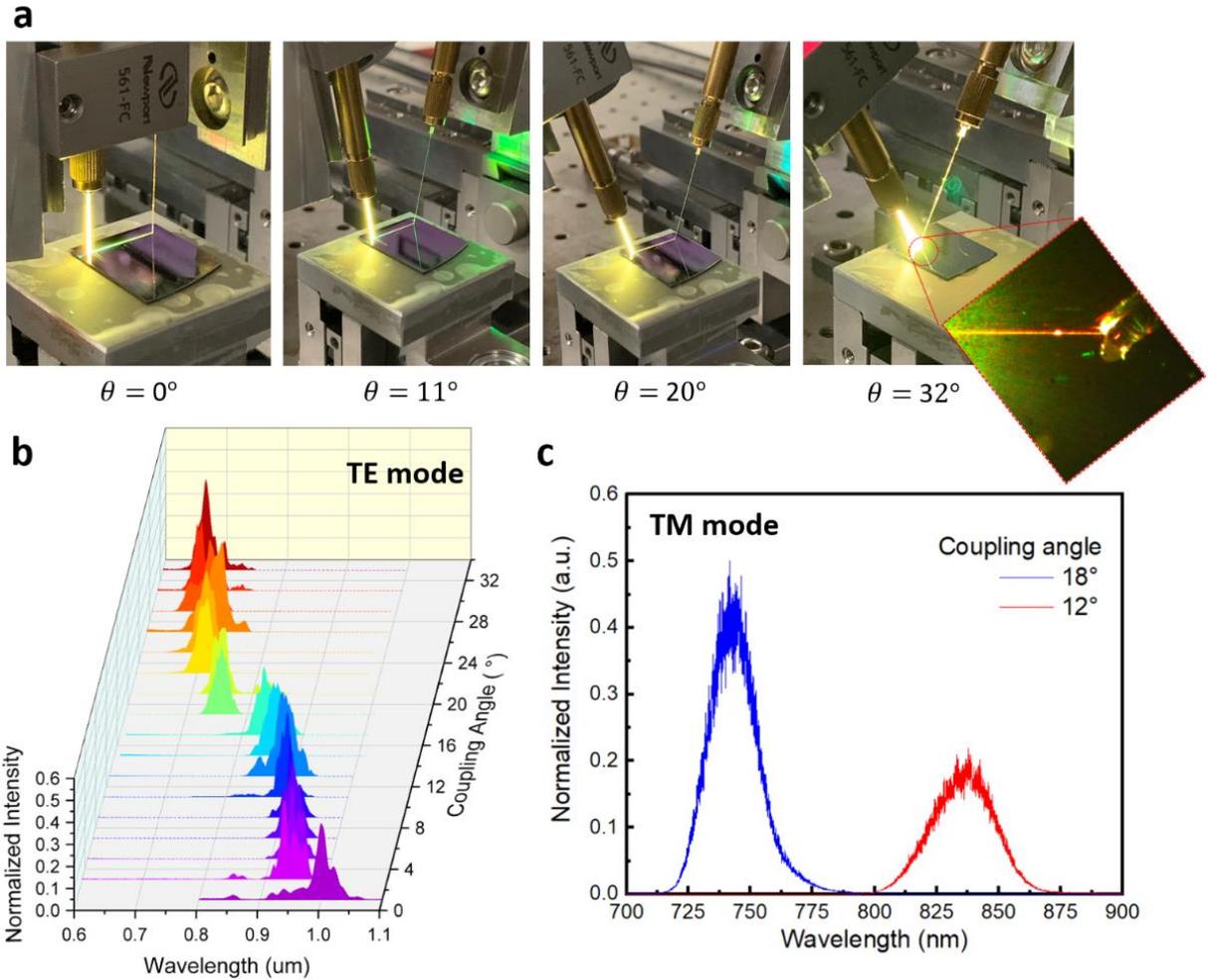

**Figure 5.** SWGC output measurement results. **a** The pictures of the fiber to SWGC coupling setup with different coupling angles; the enlarged picture shows the red-color guided light through the waveguide, coupled to the output fiber. **b** XYZ plot of the coupling efficiencies using TE mode; X-axis: wavelength, Y-axis: coupling angle, Z-axis: normalized coupling efficiency. **c** XY plot of the coupling efficiencies using TM mode; blue line: $\theta = 18°$, red line: $\theta = 12°$.

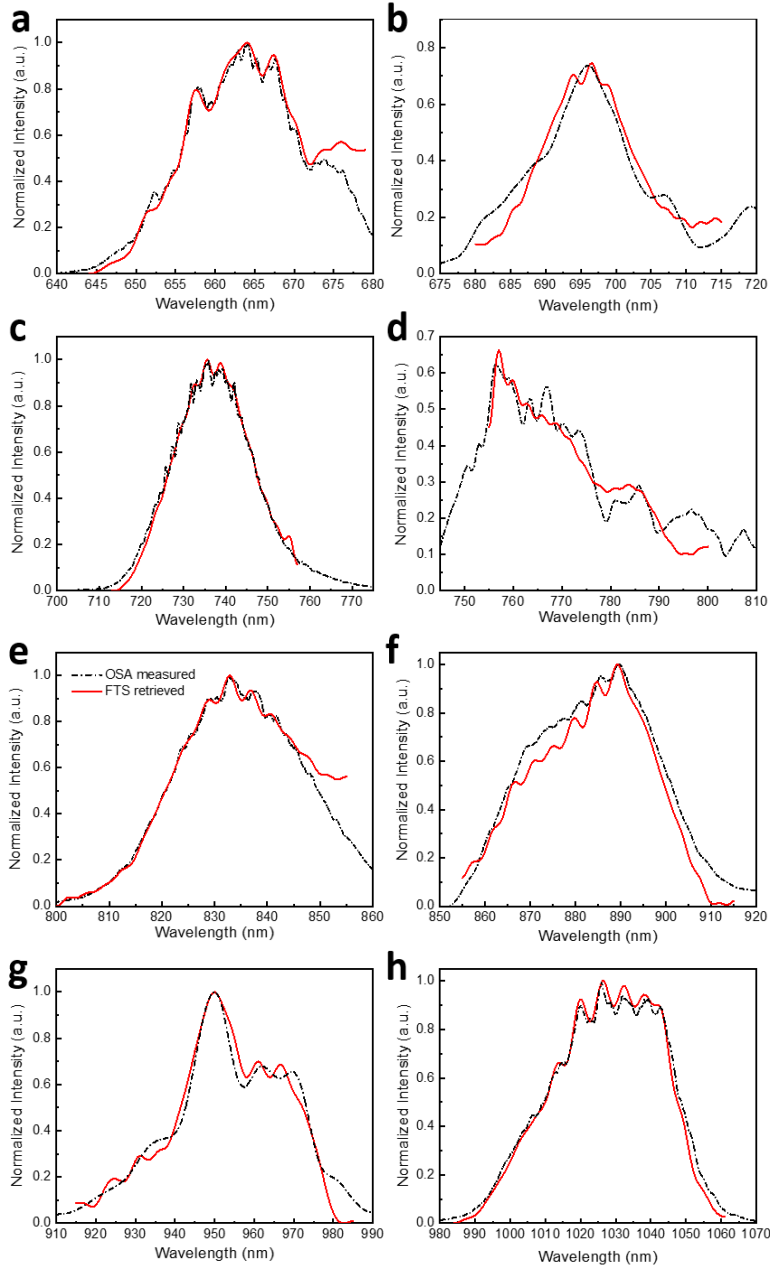

**Figure 6.** Retrieved spectra from the bandpass sampling SHFTS (red-solid lines) and direct measured spectra from the OSA (black-dotted lines) with different SWGC coupling conditions of **a** TE mode $\theta = 32°$, **b** TE mode $\theta = 25°$, **c** TM mode $\theta = 18°$, **d** TE mode $\theta = 20°$, **e** TM mode $\theta = 12°$, **f** TE mode $\theta = 14°$, **g** TE mode $\theta = 4°$, and **h** TE mode $\theta = 0°$.

# Abstract graphic

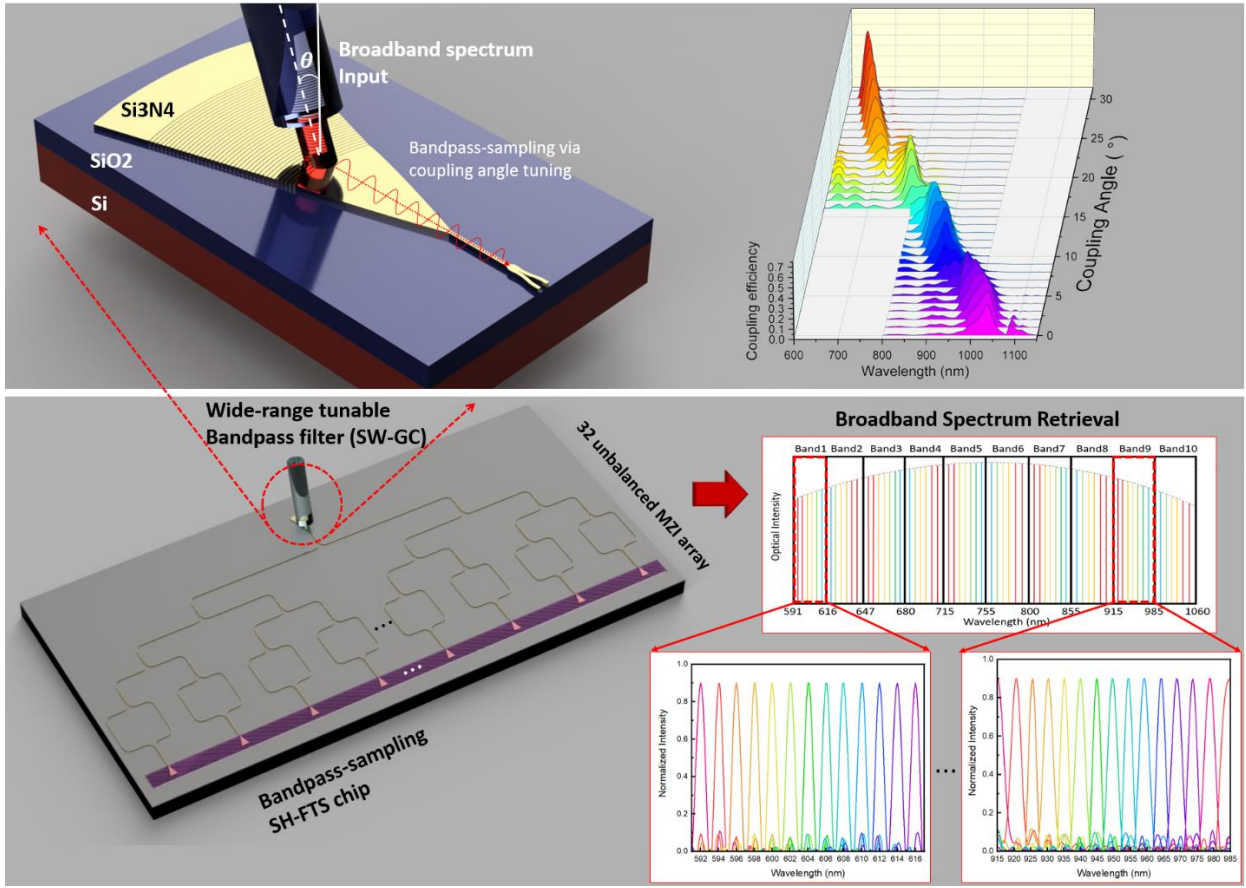

**Supporting Information**

# Dual-Polarization Bandwidth-Bridged On-Chip Bandpass Sampling Fourier Transform Spectrometer from Visible to Near-Infrared

Kyoung Min Yoo[1] and Ray T. Chen[1,2,*]

[1]Department of Electrical and Computer Engineering, The University of Texas at Austin, 10100 Burnet Rd. Austin, TX, 78758, USA.

[2]Omega Optics Inc., 8500 Shoal Creek Blvd., Bldg. 4, Suite 200, Austin, TX, 78757, USA.

*Corresponding author. E-mail: chenrt@austin.utexas.edu, yoo_eb@utexas.edu

This supplementary includes the relevant figures (Figure S1-14) to support the manuscript, and the descriptions of the strip waveguide and Y-branch design and simulation results. The sensing application results with different solutions using SHFTS are described (Figure S15-20). The video files of E-field simulations can be found in the separate files.

Video1: E-field animation of the SWGC fiber to waveguide coupling with $\theta = 3°$.

Video2: E-field animation of the SWGC fiber to waveguide coupling with $\theta = 11°$.

Video3: E-field animation of the Y-splitter at $\lambda = 900\ nm$.

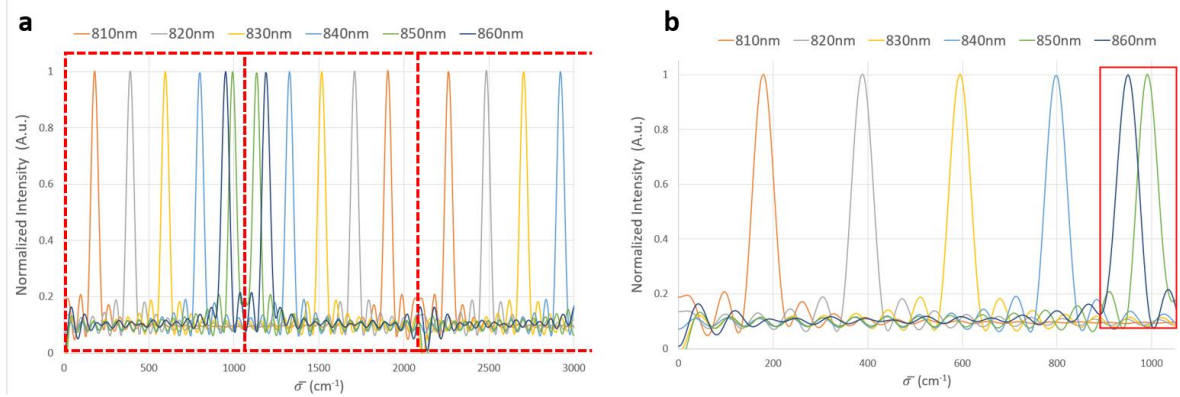

**Figure S1.** The retrieved narrow-band signals from the standard SHFTS. **a** The wavenumber-shifted copies of the original Fourier transform. **b** Zoomed-in spectrum within the bandwidth, which shows an aliasing error due to the bandwidth exceeded signal ($\lambda_o$ =860 nm).

**Silicon-nitride strip waveguide design**

The $Si_3N_4$ strip waveguide comprises of the $Si_3N_4$ waveguide core and $SiO_2$ bottom cladding on Si substrate (Fig. S2a). To ensure the broadband fundamental transverse-electric (TE0) single mode operation covering the targeting wavelength range ($\lambda = 650\ nm$ to $1050\ nm$), the width and height of the $Si_3N_4$ waveguide core are optimized as $w_{wg}$ = 500 nm and $h_{core}$ = 220 nm by scanning the effective index of the waveguide as a function of the width as shown in Fig. S2g. Using the optimized design, the effective index of guiding TE0, TM0 and TE1 modes as a function of the wavelength are shown in Fig. S2e. The result shows that the TE0 single mode covers from $\lambda$ =620 nm with the cut-off wavelength at 1030 nm. Also, to build the MZI structure with the minimum size of the footprint, the minimum radius of the 90° bend should be determined to ensure the low-loss across the whole wavelength range. By sweeping the bending radius from 1 $\mu m$, we designed the minimum bend radius of 20 $\mu m$ to guide the fundamental TE mode up to $\lambda$ =1030 nm as shown in Fig. S2f, indicating the same cut-off wavelength with the straight waveguide. Considering the propagation loss of the bent waveguide and the mode mismatching between the straight and bent waveguides, the total bending loss from 90° bend with r = 20 $\mu m$ was calcalued as 0.027 dB. The refractive index and electrical field (E-field) profiles of the optimized strip waveguide and bent waveguie are shown in Fig. S2b, c and d.

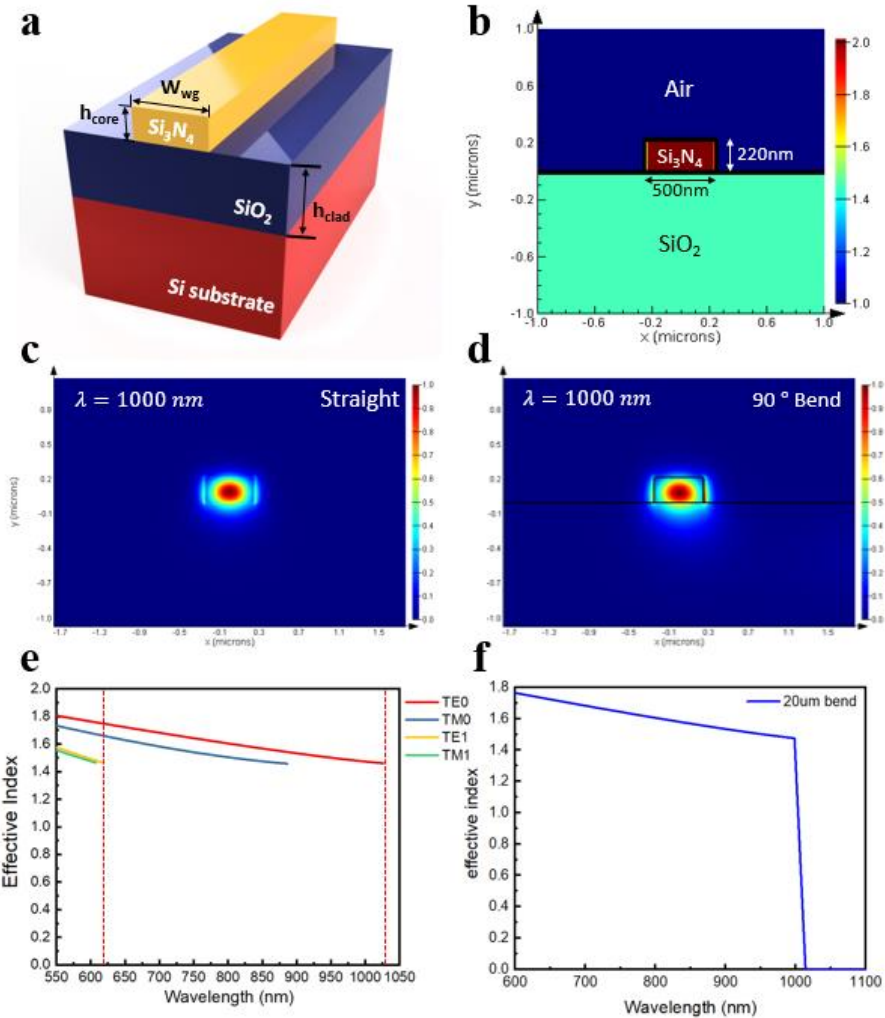

**Figure S2.** $Si_3N_4$ strip waveguide design and simulation results. **a** Schematic illustration of $Si_3N_4$ strip waveguide on $SiO_2$ bottom cladding with Si substrate. **b** Refractive index profile with $w_{wg}$=500 nm, $h_{core}$=220 nm, $h_{clad}$=2.8 $\mu m$. **c** E-field profiles of fundamental TE mode guiding through the straight waveguide and **d** 90-degree bent waveguide with r=20 $\mu m$ at $\lambda = 1000\ nm$. **e** Effective index of guiding modes in optimized strip waveguide as a function of wavelength. **f** Effective index of 90-degree bent strip waveguide with r=20 $\mu m$.

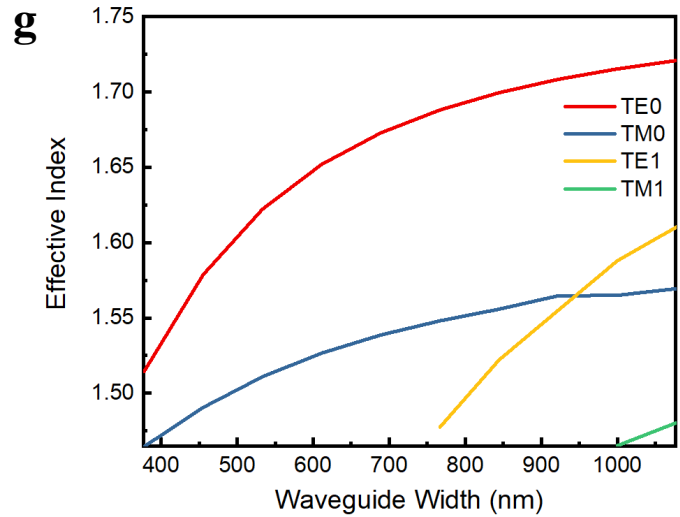

**Figure S2 (cont.). g** Effective index of the Si$_3$N$_4$ strip waveguide as a function of w$_{wg}$ at $\lambda = 800\ nm$.

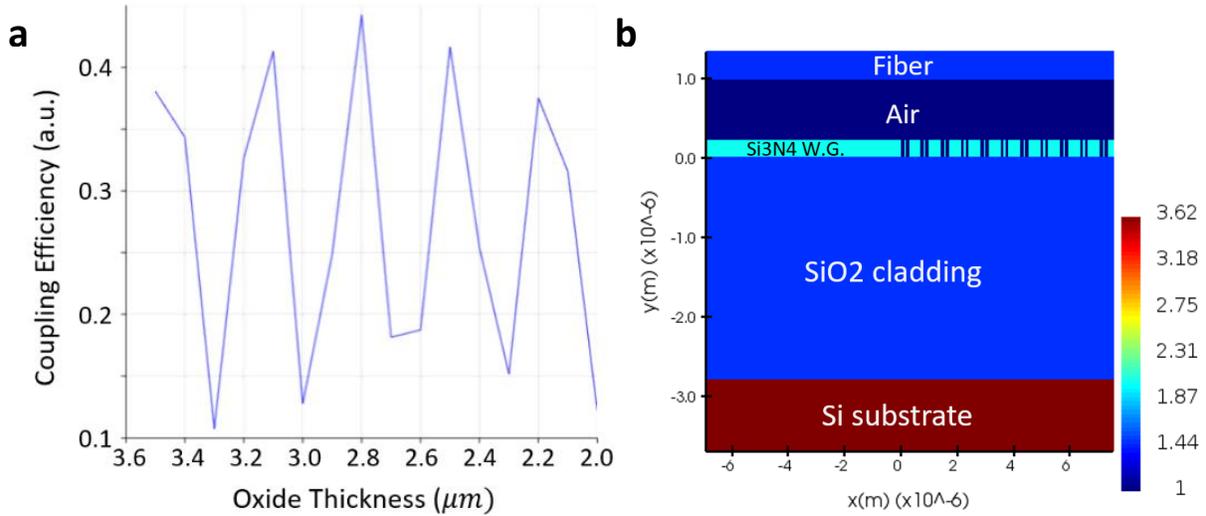

**Figure S3. a** SWGC fundamental TE mode coupling efficiencies at $\lambda = 900\ nm$ as a function of the thickness of the SiO$_2$ bottom cladding. **b** Refractive index profile of the optimized SWGC structure.

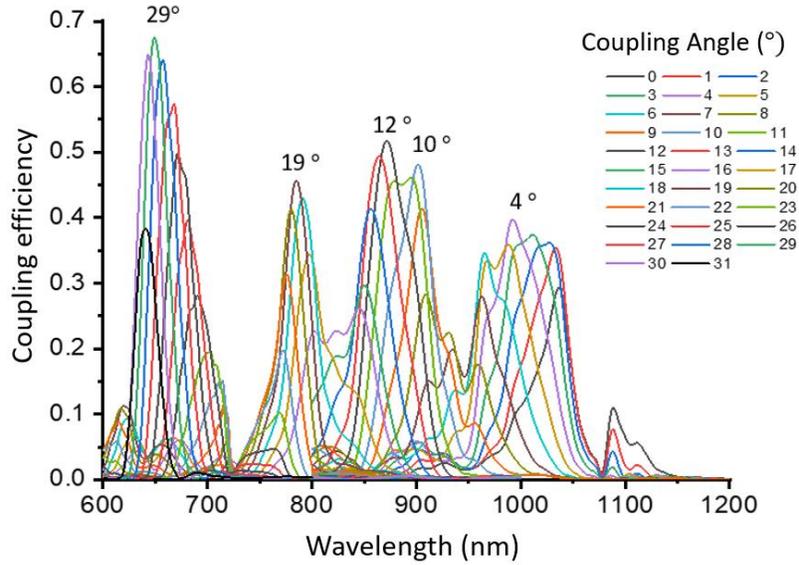

**Figure S4.** XY plot of the coupling efficiencies of the SWGC with different coupling angle from 0° to 31° as a function of wavelength.

### Y-branch splitter and combiner design

Y-branch splitter and combiner are essential components to build MZI structure to equally divide and combine the light with 50:50 ratio. We optimized the Y-branch to minimize the loss using Lumerical Mode and FDTD simulations; the width of the splitter body is divided into 5 parts ($w_{wg}$, $w_1$, $w_2$, $w_3$, $w_{gap}$) and optimized the dimensions by the built-in particle-swarm algorithm to achieve the minimum loss for the fundamental TE mode across the wavelength range covering from 600nm to 1000 nm. The final optimized dimensions are shown in Table S1.

|  | $W_{wg}$ | $W_1$ | $W_2$ | $W_3$ | $W_{gap}$ |
|---|---|---|---|---|---|
| **Width (nm)** | 500 | 1128 | 1395 | 1360 | 200 |

**Table S1.** Y-branch optimized dimensions

The schematic illustration of Y-branch structure and E-field simulation result with optimized dimensions are shown in Fig. S5a and b. To analyze the power-splitting performance, the S-parameters of output ports were calcultated (S21, S31) across $\lambda = 600 - 1000\ nm$. Fig. S5c shows the transmitted light intensites of output ports (port 2 and 3) with the normalized intensity of 1 from the input port (port 1); in other words, if there is no loss from the Y-splitter, each output port should have 0.5 of transmitted power. Using the optimized structure, we could get the minimum loss of 0.5 dB from each output ports at $\lambda = 800 \ nm$. Furthermore, the same Y-branch structure can be used as a combiner in opposite direction. Fig. S5d shows the gradiant map of output power as functions of two input powers. Using the optimized structure, we could achieve ~1 dB combining loss when there are two equal inputs of the same phase. Also, we checked the TM mode compatiblity using the same design, and it turned out that the loss is higher (~2 dB) than that of the TE mode operation as shown in Fig. S5e. The video animation of the top-view E-field simulation of the Y-splitter is attached in the supporting information.

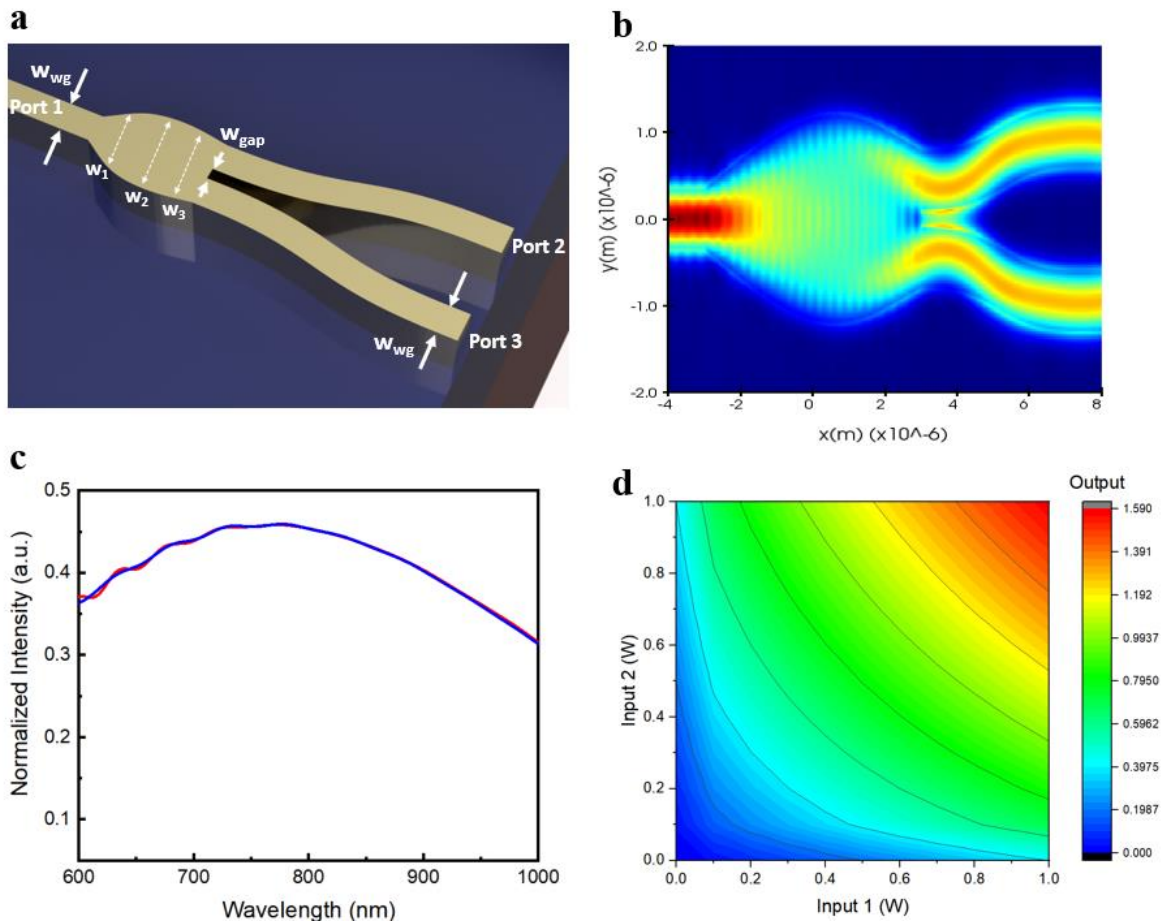

**Figure S5. a** 3D schematic illustration of the Si$_3$N$_4$ Y-branch design. **b** E-field of optimized Y-branch design; Input port: port 1, Output ports: port 2 and 3. **c** Transmitted light intensities of the fundamental TE mode from port 1 to port 2 and 3 (S21 and S31) of Y-splitter. **d** Output power (port1) gradient map as a function of input powers (x-axis: port2, y-axis: port 3) of the Y-combiner.

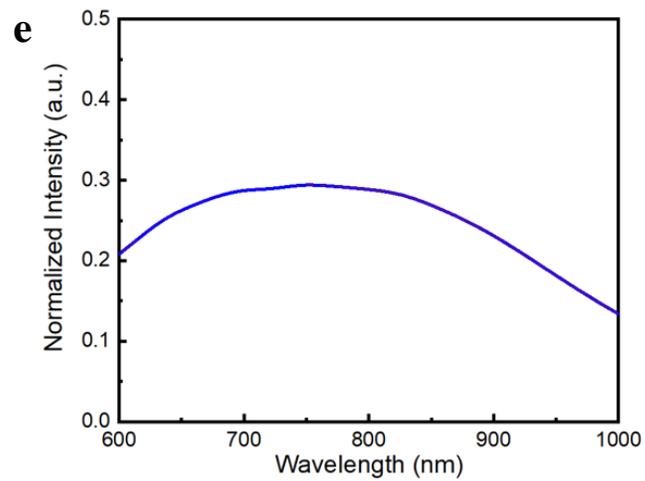

**Figure S5 (cont.). e** Transmitted light intensities of the fundamental TM mode as a function of wavelength measured from two output ports.

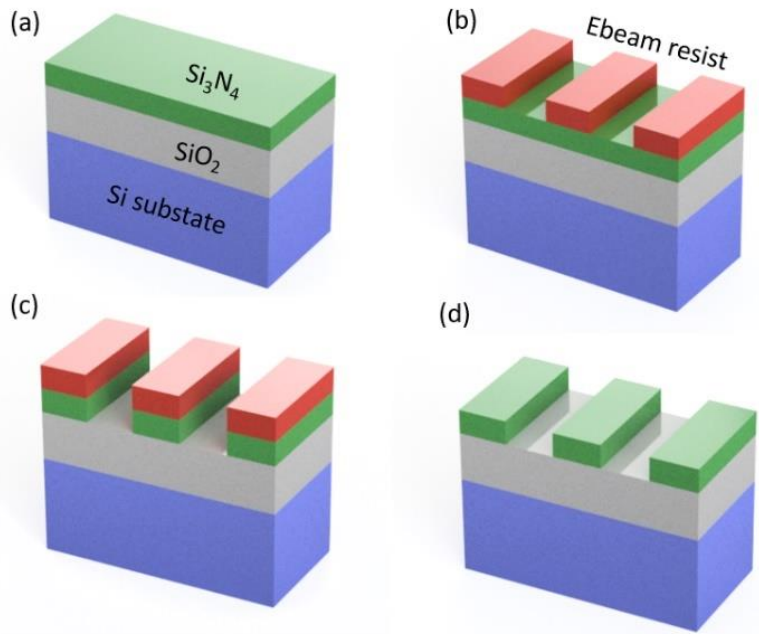

**Figure S6.** Schematic illustration of the fabrication process. **a** Si3N4-on-SiO2 wafer. **b** Deposit 400 nm E-beam resist (ZEP-520A) for e-beam lithography. **c** Etch Si3N4 layer with RIE. **d** Post-fabrication treatment with removal PG.

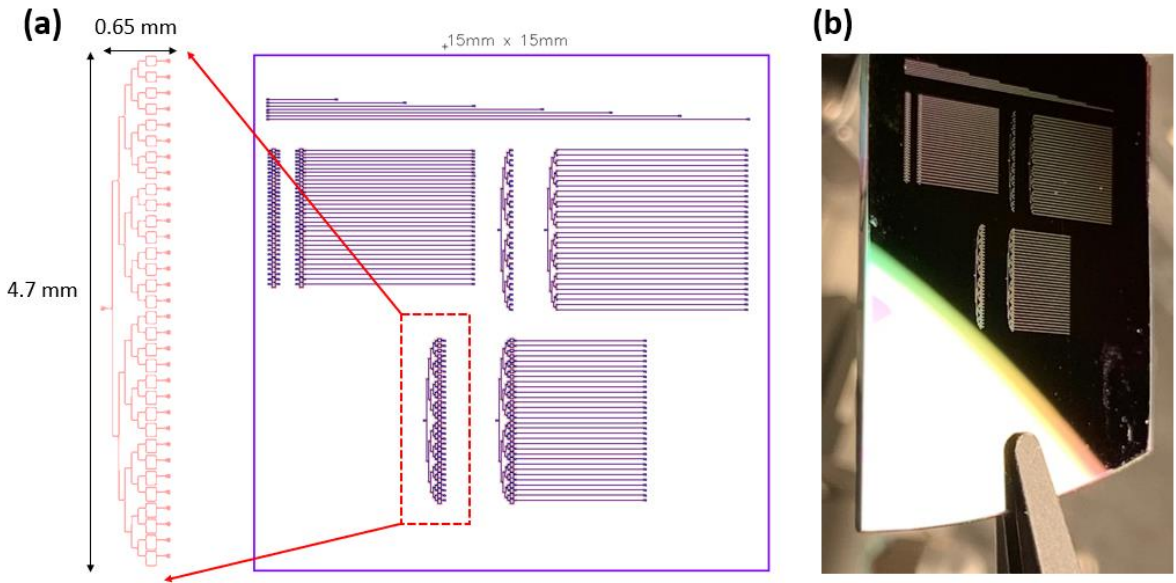

**Figure S7.** Bandpass sampling SHFTS device footprint and optical images. **a** The device footprint including characterization patterns. The size of SHFTS is around 4.7mm x 0.65 mm. **b** The picture of fabricated Si3N4 chip.

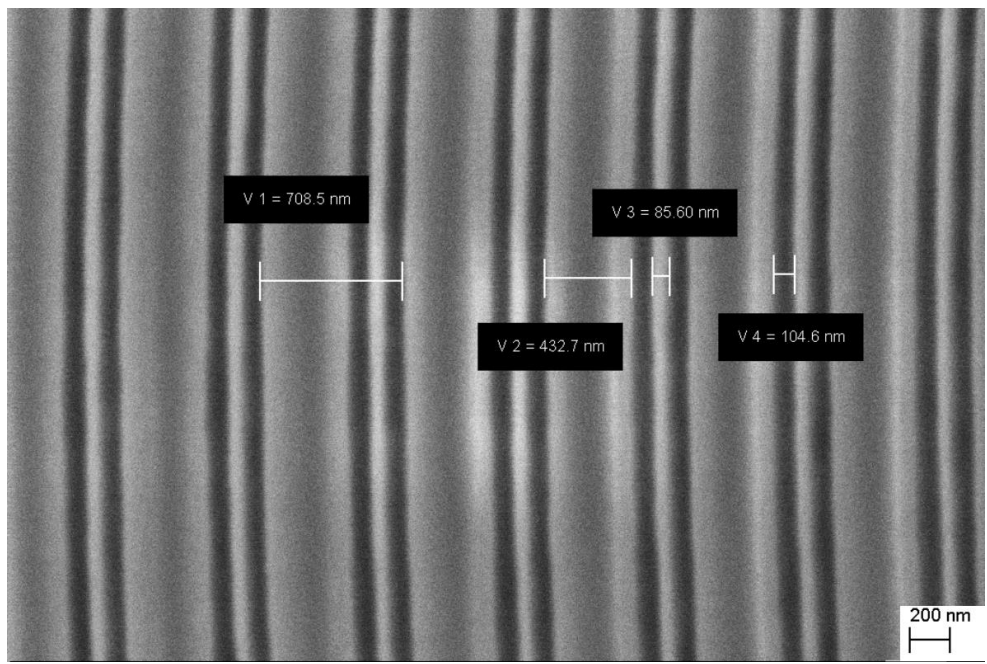

**Figure S8.** Zoomed-in SEM image of SWG structures in SWGC.

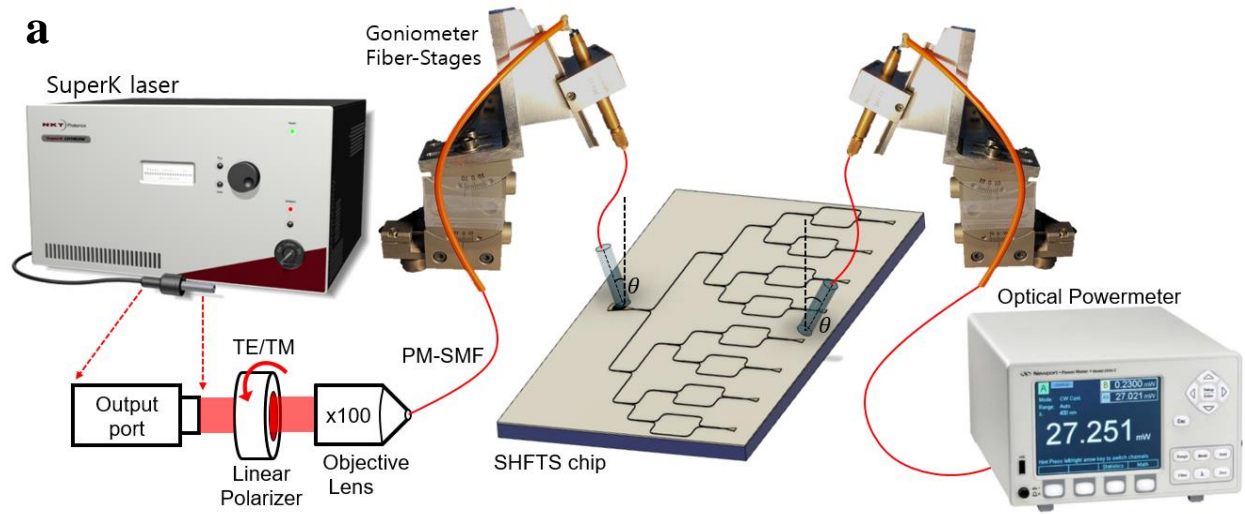

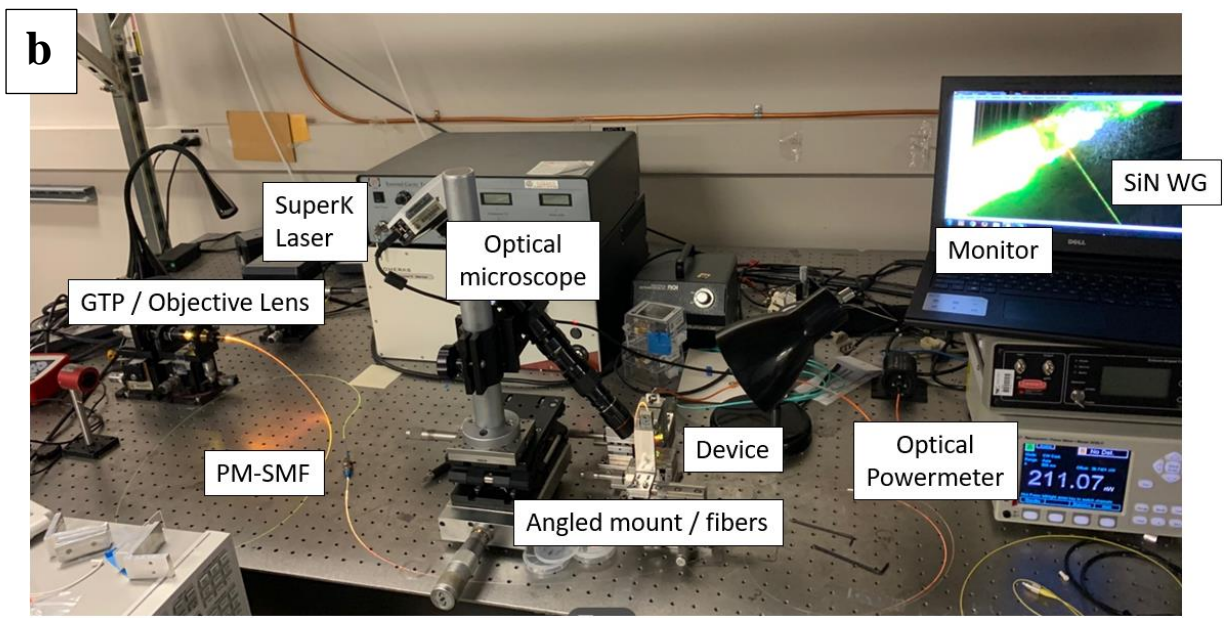

**Figure S9. a** A schematic diagram of the broadband SuperK laser measurement setup and **b** the picture of broadband SuperK laser measurement setup; GTP: Glan-Thompson Linear Polarizer, PM-SMF: polarization-maintaining single mode fiber.

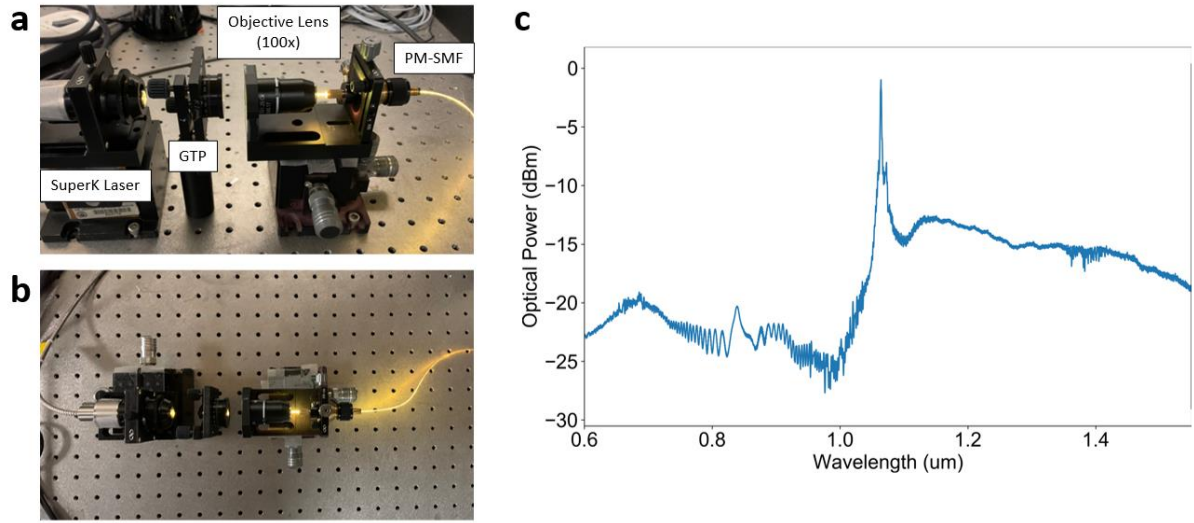

**Figure S10.** The pictures of the SuperK laser coupling setup with **a** side- and **b** top-view. **c** Optical spectrum of the broadband SuperK laser coupled to the PM-SMF measured by OSA; GTP: Glan-Thompson linear polarizer, PM-SMF: polarization-maintaining single mode fiber, OSA: optical spectrum analyzer.

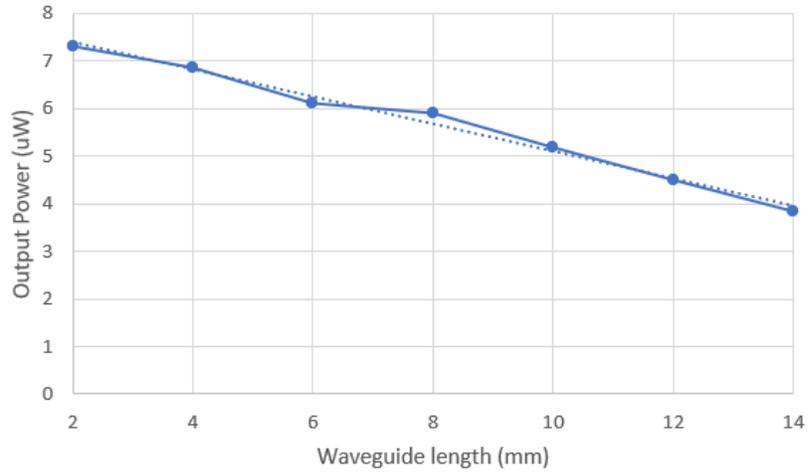

**Figure S11.** The output powers from the $Si_3N_4$ strip waveguides with various lengths for propagation loss measurement.

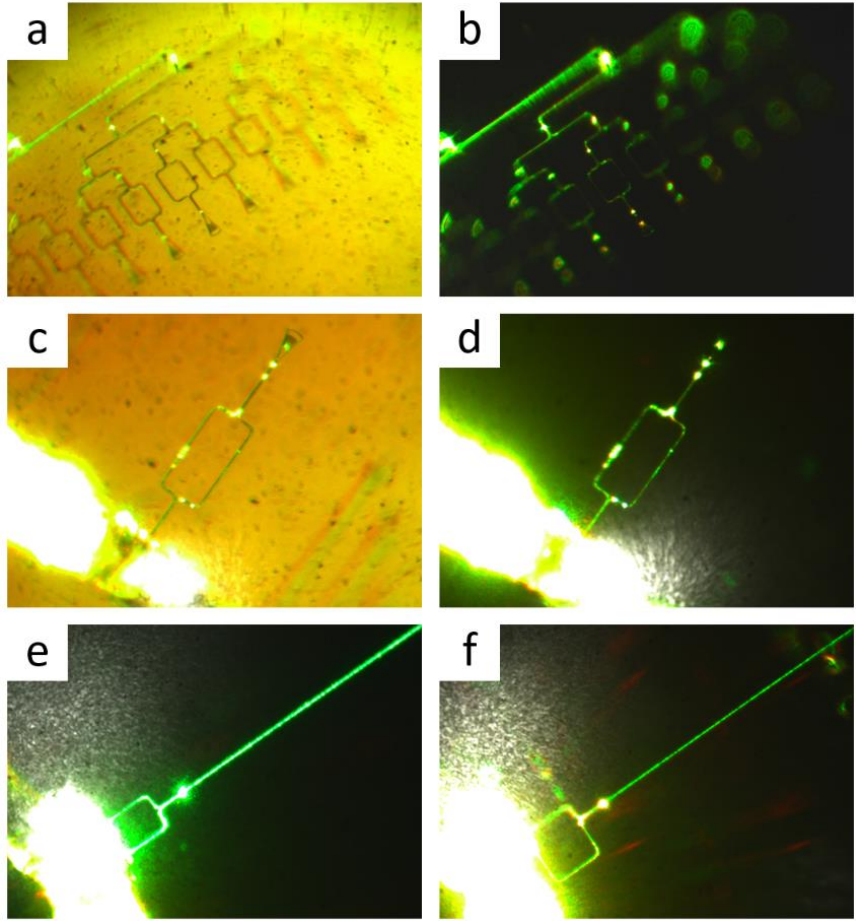

**Figure S12.** Optical microscope images of light coupled SHFTS and MZI structures with background light on/off. **(a-b)** SHFTS, **(c-f)** MZIs.

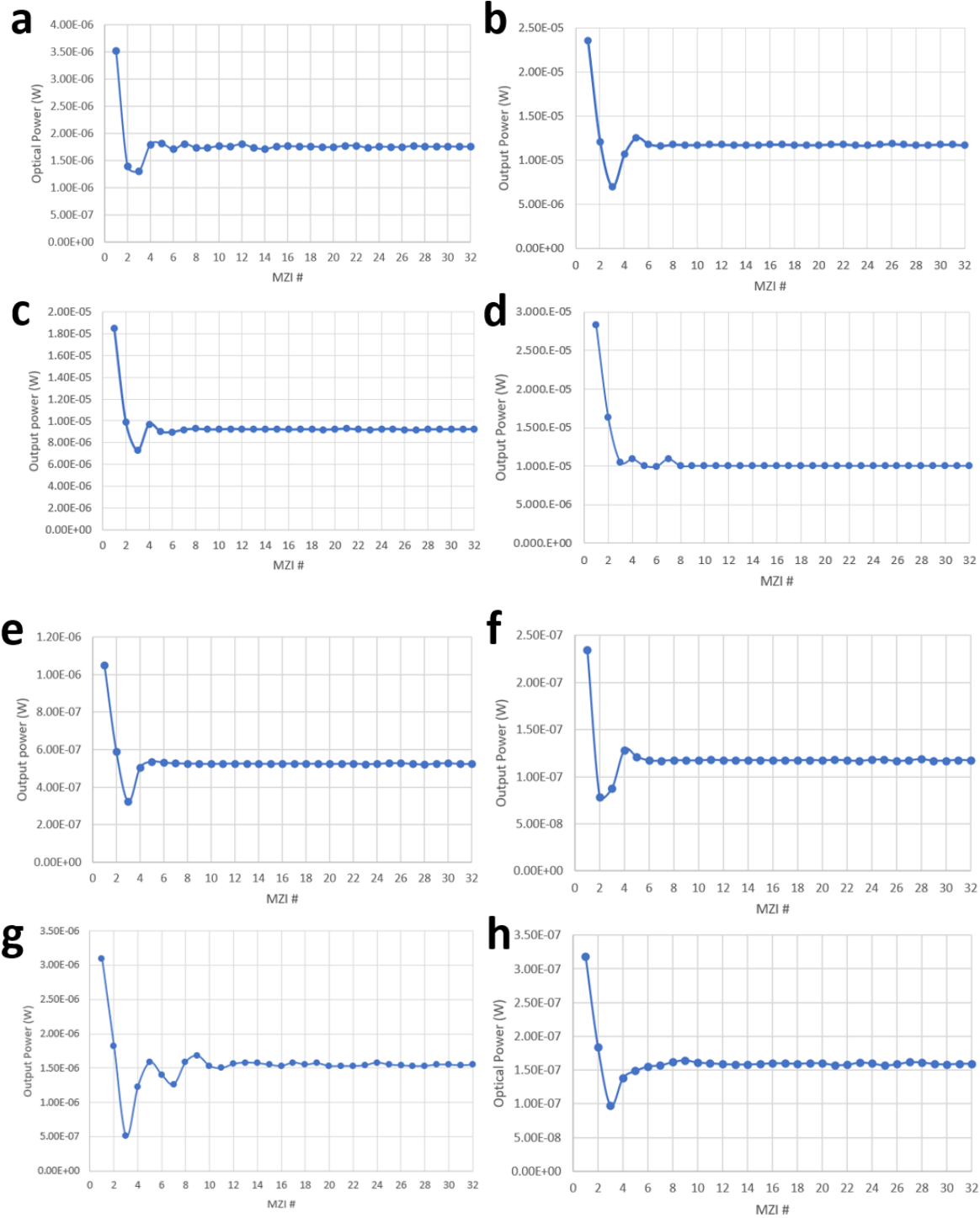

**Figure S13.** Output power measurement results using bandpass sampling SHFTS with different SWGC coupling conditions of **a** TE mode $\theta = 32°$, **b** TE mode $\theta = 25°$, **c** TM mode $\theta = 18°$, **d**

TE mode $\theta = 20°$, **e** TM mode $\theta = 12°$, **f** TE mode $\theta = 14°$, **g** TE mode $\theta = 4°$, and **h** TE mode $\theta = 0°$.

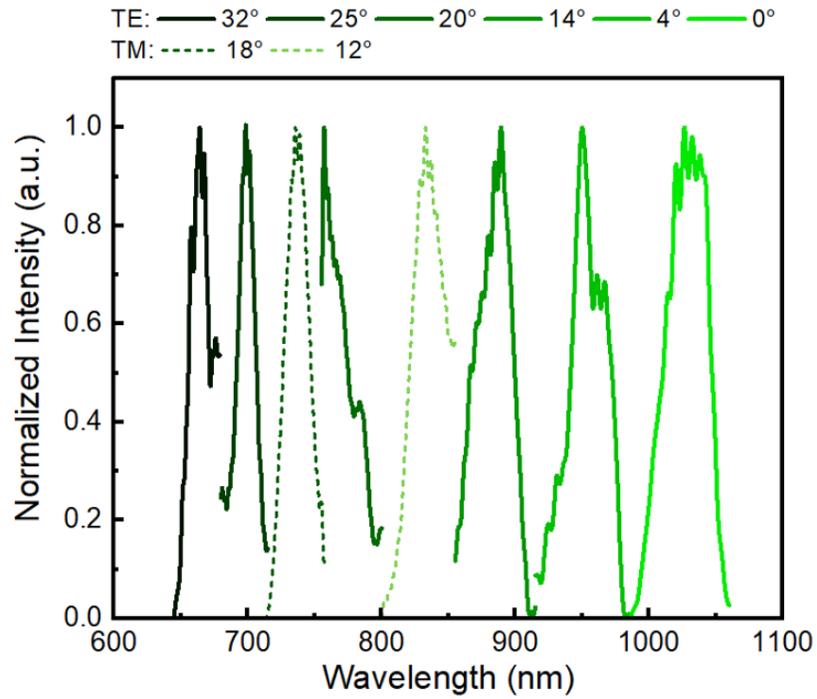

**Figure S14.** Broadband spectrum retrieval using bandpass sampling SHFTS by changing the grating coupling angles and polarizations (TE and TM).